\documentclass[useAMS,usegraphicx,usenatbib]{mn2e}
\usepackage{graphicx}
\usepackage{aas_macros}     % for journal abbreviations in references
\usepackage{amsmath}
\usepackage[usenames,dvipsnames]{xcolor}

\usepackage{fixltx2e}[1999/12/01]

\newcommand{\Mass}{\ensuremath{ h^{-1} M_{\odot}}}
\newcommand{\Mpc}{ \ensuremath{h^{-1} {\rm Mpc}} }
\newcommand{\Gpc}{ \ensuremath{h^{-1} {\rm Gpc}} }

\newcommand{\Mstar}{ \ensuremath{ M_{\rm star}} }
\newcommand{\Mhalo}{ \ensuremath{ M_{\rm halo}} }

\newcommand{\vect}[1]{\bmath{#1}}

\title[Galaxy formation on the largest scales]
{Galaxy formation on the largest scales: The impact of astrophysics
on the BAO peak}

\begin{document}
\setlength{\topmargin}{-1.5cm}

\author[Angulo et al]{
\parbox[h]{\textwidth}
{R. E. Angulo$^{1,2,3}$ \thanks{rangulo@stanford.edu}, S. D. M. White$^{3}$, V. Springel$^{4,5}$, B. Henriques$^{3}$} \vspace*{6pt} 
\\  $^{1}$Centro de Estudios de F\'isica del Cosmos de Arag\'on, Plaza San Juan 1,  Planta-2, 44001, Teruel, Spain.
\\ $^2$  Kavli Institute for Particle Astrophysics and Cosmology,\\ 
Stanford University, SLAC National Accelerator Laboratory, Menlo Park, CA 94025, USA
\\ $^3$ Max-Planck-Institute for Astrophysics, Karl-Schwarzschild-Str. 1, 85740 Garching, Germany.
\\ $^4$ Heidelberg Institute for Theoretical Studies, Schloss-Wolfsbrunnenweg 35, 69118, Heidelberg, Germany,
\\ $^5$ Zentrum f\"{u}r Astronomie der Universit\"{a}t Heidelberg, ARI, M\"onchhofstr. 12-14, 69120 Heidelberg,
Germany.
}
\maketitle

\date{\today}
\pagerange{\pageref{firstpage}--\pageref{lastpage}} \pubyear{2011}
\label{firstpage}

\begin{abstract} 
We investigate the effects of galaxy formation on the baryonic
acoustic oscillations (BAO) peak by applying semi-analytic modelling
techniques to the Millennium-XXL, a $3 \times 10^{11}$ particle N-body
simulation of similar volume to the future EUCLID survey. Our approach
explicitly incorporates the effects of tidal fields and stochasticity
on halo formation, as well as the presence of velocity bias, spatially
correlated merger histories, and the connection of all these with the
observable and physical properties of galaxies. We measure significant
deviations in the shape of the BAO peak from the expectations of a
linear bias model built on top of the nonlinear dark matter
distribution. We find that the galaxy correlation function shows an
excess close to the maximum of the BAO peak ($r\sim110\Mpc$) and a
deficit at $r\sim90\Mpc$.  Depending on the redshift, selection
criteria and number density of the galaxy samples, these bias
distortions can be up to 5\% in amplitude. They are, however, largely
absorbed by marginalization over nuisance parameters in current
analytical modelling of the BAO peak in configuration space, in particular into
the parameter that controls the broadening due to nonlinear evolution. As a
result, the galaxy formation effects detected here are unlikely to bias the
high-precision measurements planned by the upcoming generation of wide-field
galaxy surveys.
\end{abstract}

\section{Introduction}

The simple $\Lambda$CDM cosmological model has been supported
consistently over the last $20$ years by virtually every observational
probe.  Measurements of the cosmic microwave background (CMB),
large-scale structure (LSS), type-Ia supernovae, the Ly-$\alpha$
forest, weak gravitational lensing, and the abundance of galaxy
clusters, all seem to point towards the existence of dark energy and
dark matter, with a cosmology based on ordinary general relativity and
Gaussian primordial density fluctuations \citep[e.g.][]{Komatsu2011,
  Sanchez2012, Beutler2012, Suzuki2012,Fu2008,Rozo2010,
  PlanckCosmo2013, Viel2013}. This overwhelming and diverse
observational evidence is, nevertheless, still not fully conclusive,
and the lack of any convincing independent indication of the nature of
either dark matter or dark energy has fueled many observational
campaigns to provide more precise tests further test of $\Lambda$CDM
predictions (e.g. DES, HEDTEX, J-PAS, MS-DESI, CHIME, TAIPAN, Euclid,
LSST). The discovery of any departure from the vanilla
$\Lambda$CDM model would help us to understand better the dominant
constituents of our Universe.

One of the simplest astrophysical observations -- measuring the
redshift and angular position of galaxies on the sky -- provides one
of the most powerful ways to constrain the cosmological
model. Statistical measurements of the spatial distribution of
galaxies can be compared with theoretical models and, in this way, can
be used to derive constraints on the parameters of a given
cosmological model. This exercise is most commonly done on large
scales \citep[but see][]{Cacciato2013, Simha2013}. In particular, some
of the strongest current constraints on the expansion history arise
from measurements of the position of a particular feature of scale
$\sim110\Mpc$: the Baryonic Acoustic Oscillation (BAO) peak
\citep{Eisenstein2005,Cole2005,blake_wigglez_2011,beutler_6df_2011,
  anderson_clustering_2012}.

A fundamental assumption underpinning the extraction of cosmological
information from such measurements is that, on large scales, the complex
astrophysical processes responsible for galaxy formation can be
decoupled from gravity and represented by a set of nuisance
parameters, thus allowing the galaxy abundance in a given volume element
to be related directly to the underlying primordial mass density,
which, in turn, can be described using linear perturbation
theory. In this way, a working description of the observables can be built
and cosmological constraints can be derived.

This simplified picture may, however, be too crude to obtain precise
results.  The underlying dark matter (DM) density field differs
nontrivially from that expected in linear theory. Nonlinear effects
due to mode coupling weaken and distort the BAO peak.  Nonlinear
evolution also breaks the simple relation between density and velocity
divergence predicted by linear theory, with implications for the
measurement of clustering from redshifts which reflect both the Hubble
flow and the peculiar velocities of galaxies: these so-called
``redshift space distortions'' (RSD) further weaken the BAO peak.
Finally, the relation between the mass density field and the galaxy
distribution is complex.  In particular, galaxies trace the underlying
field in a biased manner because they form at the bottom of local
potential wells (the centres of DM haloes) not at the locations of
random DM elements. This may imply that they trace the underlying
velocity field in a special way too, introducing velocity biases.
Moreover, halo formation is a function not only of the local DM
density, but also of the tidal field and other aspects of the local
environment.  Finally, the relation between DM haloes and the galaxies
they contain is not set at their observed redshift, but rather
reflects their entire assembly history. This has particularly marked
effects on satellite galaxies.

The effects of nonlinear evolution and redshift space distortion on the BAO
signal in the DM and halo distributions have been studied and quantified
extensively in recent years
\citep{Seo2003,Angulo2005,Seo2005,Angulo2008,Smith2008,Sanchez2008,Seo2010}.
They have been shown to shift the location of the BAO peak and, if uncorrected,
to bias estimates of cosmological parameters. In comparison, the impact of
galaxy formation has been much less explored. This reflects the difficulty in
estimating its effects theoretically.  State-of-the-art $N$-body simulations
are just beginning to achieve the volume and mass resolution required to follow
the assembly of the haloes expected to host typical galaxies over the volumes
to be targeted by upcoming surveys \citep{Kim2011,Angulo2012a,Watson2013}. {\it
Ab initio} hydrodynamical simulations of galaxy formation over the required
volumes are still far beyond current computational capabilities.

Our limited understanding of the effects of galaxy formation on the BAO signal
contrasts with the importance of the topic. The extraction of robust and
precise information about the dark energy from future galaxy surveys will rely
critically on a quantitative understanding of this issue.

In this paper, we perform the most realistic modelling of the galaxy population
on large scales to date, explicitly showing that BAO measurements in
configuration and redshift space can indeed be affected by galaxy formation
physics.  We do this by combining a large DM only $N$-body simulation with a
post-processing scheme for simulating galaxy formation. The first of these
ingredients follows self-consistently the nonlinear growth of structure, the
formation of self-bound objects, their connection to the velocity and tidal
fields, and their evolution over cosmic time. The second couples a collection
of physically motivated equations capturing the many processes relevant for
galaxy formation to the accretion, merger history and dynamics of the DM halos
and subhalos. We note that this procedure explicitly takes into account the
assembly history of each DM halo and its correlation with the surrounding
large-scale overdensity field.

These tools allow us to show explicitly that galaxy clustering on very large
scales, and the BAO peak in particular, is not a linearly scaled version of DM
clustering, as is often assumed. Deviations from this model are small, but will
nevertheless have to be understood quantitatively in order to fully exploit
future clustering measurements, and are thus a requirement if they are to
fulfil their promise as robust and precise probes of the dark energy equation
of state and the nature of the gravity law.

The layout of this paper is the following. First, in Section~2.1 we present the
$N$-body simulation used for this work, and in Section~2.2 we give a brief
summary of the main features of our semi-analytic model for galaxy formation.
In Section~2.3 we describe the implementation of this model on our $N$-body
simulation and how we overcame problems due to its relatively poor mass
resolution. We present our main results in Section~3 and 4, focusing first on
the connection between individual haloes and their galaxies, and then on an
exploration of clustering statistics. Finally, we conclude in Section~5.

\section{Methods}

\subsection{The MXXL $N$-body simulation}

The $N$-body simulation used for this paper, Millennium-XXL (MXXL), is the
latest member of the Millennium Simulation series \citep[MS,
MS-II]{Springel2005a,Boylan-Kolchin2009}. Extensive details of the simulation
were given in \cite{Angulo2012a,Angulo2012b}. Here we just provide a short
description of the main features relevant for this work.

The MXXL uses $6720^3$ particles, each of mass $m_p = 6.1\times10^{9}\,\Mass$,
to simulate the growth of cosmic structure over a volume of $ V =
27\,(\Gpc)^3$. This combination of mass resolution and volume is adequate to
obtain good statistical precision on the scales relevant for BAO and RSD
measurements, while resolving the halos and subhalos expected to host the
galaxies targeted by next-generation, large-volume surveys.  The cosmological
parameters and output times were set to match those of the two previous
Millennium simulations, specifically, $h=0.73$, $\Omega_m=0.25$,
$\Omega_{\Lambda}=0.75$, $n=1$ and $\sigma_8=0.9$. Halo/subhalo catalogues are
stored for 63 snapshots, roughly spaced by $300\,{\rm Myr}$ at $z\sim0$. The
cosmological parameters differ from those preferred by more recent cosmological
analyses \citep[see][for a way to adjust for this]{Angulo2010a}, however, this
has little importance for predicting the spatial distribution of the galaxy
population as shown by \cite{Guo2013}. The initial conditions were generated at
$z=63$, which ensures that artificial transients are negligible for
low-redshift clustering statistics \citep{Angulo2008}. 

At each snapshot, haloes with more than $20$ particles were identified by a FoF
algorithm \citep{Davis1985}, and self-bound subhaloes with more than $15$
particles were found using {\tt SubFind} \citep{Springel2005a}. Both of these
operations were performed on-the-fly during the $N$-body calculation, which
reduced significantly the overall I/O workload and disk usage.

Subhalo merger trees were built subsequently by identifying, for each subhalo
in each snapshot, the most likely descendant in the following snapshot. This
was done by tracking the $15$ most bound particles of the subhalo, giving them
a weight proportional to $j^{-2/3}$, where $j$ is the particle position in a
list ordered by binding energy. This approach gives more importance to the
particles expected to best represent the orbit of hypothetical galaxies. We
note that our tree-building scheme is slightly simpler than that employed for
the MS (for instance, we do not perform backwards checks on the descendants,
nor look for descendants more than one snapshot ahead). We have, however,
explicitly checked that this has no impact on the quantities explored in this
paper.

Using these procedures, we constructed $600$ million distinct trees with a root
halo at $z=0$, containing a total of over $25$ billion nodes. These structures
are the backbone of our modelling of the galaxy population in the MXXL.

\subsection{The semi-analytic model of galaxy formation}

In this paper, the properties of galaxies within our simulated volume will be
generated using the semi-analytic galaxy formation code {\tt L-Galaxies}
\citep{Springel2005a}. This code couples the (sub)halo merger trees described
in the previous subsection (from which the mass growth, dynamics and spatial
distribution are taken), with a system of differential equations that encode
the key physical mechanisms for galaxy formation. In particular, processes such
as gas cooling, star formation, feedback from SN and AGN, galaxy mergers, black
hole formation and growth, and generation of metals are all implemented in a
self-consistent manner. The philosophy and methods of semi-analytic models can
be found in the review of \cite{Baugh2006}, whereas specific aspects of recent
versions of {\tt L-Galaxies} are described in full detail by \cite{Guo2011},
\cite{Henriques2012} and references therein. Here, we simply note that this model
successfully reproduces many (though certainly not all) observable properties
of high and low-$z$ galaxies.

Most important for this paper is the fact that our framework is
physically consistent. All galaxy formation takes place at the centres
of DM subhalos.  Further, the galaxy population of a given DM halo
does not depend on its mass alone, as assumed in most Halo Occupation
Distribution (HOD) models, but also on the details of the halo's
assembly history. Thus, a recently-formed halo will typically contain
a central galaxy with younger stars than a halo of similar mass which
formed earlier, and galaxies with strong AGN activity are typically
found in more massive halos than galaxies of similar stellar mass
which are not active. (This is because feedback from the black hole
effectively decouples galaxy growth from halo growth in the active
case.)  Such processes connect the large-scale environment with galaxy
properties and may produce observable effects in wide-field galaxy
surveys.

The complex relation between DM and galaxies implies that a galaxy sample
selected according to specific observable properties will not, in general, be
simply a scaled version of the underlying DM field, nor even correspond to the
halo distribution weighted by an average mass-dependent occupation number.
Instead, such samples contain non-trivial correlations between scales, and
between the mass distributions at different times. These may be manifest, for
instance, as an ``assembly'' bias
\citep{Gao2005,Wechsler2006,Gao2007,Croton2007,Angulo2008b} or as a distortion
in the appearance of the large-scale structure \citep[e.g.][]{Bower1993}. These
possibilities demonstrate the importance of a realistic model for galaxy
formation, capable of quantifying how such effects bias measurements of
cosmological parameters from galaxy clustering observations.

\subsection{Implementation}

%==================================================================
\begin{figure} 
\includegraphics[width=\linewidth]{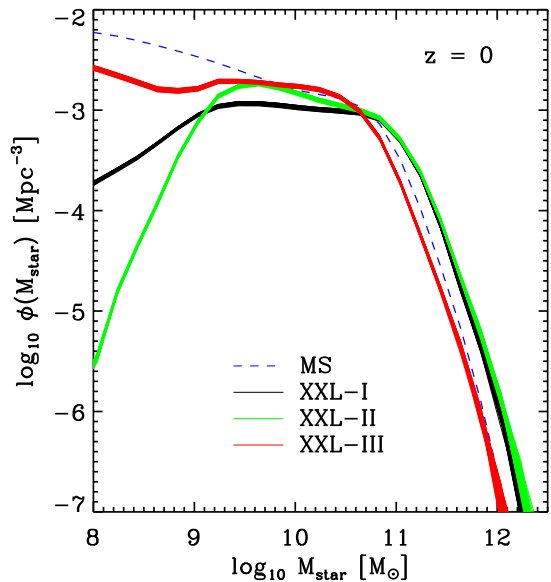}
\caption{The present-day stellar mass function predicted by a
  semi-analytic code for galaxy formation applied to dark matter only
  $N$-body simulations. The dashed line shows the result for the MS,
  while solid lines show three predictions for the MXXL: The black
  line is based on the original L-Galaxies code, without any
  corrections for the finite resolution of the MXXL. For the green
  line, each leaf in our merger trees was initialised with MS galaxies
  from a halo of similar mass and redshift. For the final MXXL model
  (shown in red) we also included (statistically) mergers which are
  unresolved in the MXXL.  Note that these modifications progressively
  improve the agreement between the MS and the MXXL. In the MXXL case,
  we display $216$ different curves, each representing a a cubic
  subregion with the same volume as the MS.
\label{fig:mf} }
\end{figure}
%==================================================================

The MXXL is the highest resolution simulation of the Universe on
very large scales, yet its mass resolution is not high enough for
the semi-analytic code to produce results convergent with those of the
higher resolution MS/MS-II. This is true even for the most massive
galaxies expected in our simulation volume. The disagreement can be
seen by comparing the black solid and blue dashed lines in
Fig.~\ref{fig:mf}, which shows $\phi$, the number of galaxies per
logarithmic stellar mass bin and per unit volume. The relatively small
difference in mass resolution (a factor of $\sim7$) allows the MS to
model galaxies reliably to a hundred times lower mass than in the
MXXL -- this is a consequence of the steep relation between halo mass,
$M_{200}$, and stellar mass, $M_{\rm star}$, over the relevant
range of halo masses. Compared to the MS, galaxies in the
MXXL are less massive below $M^*$ (the knee of the stellar mass
function), and more massive above $M^*$.

There are two main reasons for these differences in stellar mass
function shape. Both stem from the fact that the minimum halo mass
resolved by our simulation is $1.22\times10^{10}\,\Mass$, which means
that halos of this mass almost never have a resolved progenitor in the
MXXL merger trees. Although not shown here, we explicitly checked the
fraction of haloes with no progenitors (``leaves'') in the MXXL, as a
function of halo mass and time. At high redshift, virtually every halo
below $10^{11}\,\Mass$ is a leaf. At lower redshifts, the fraction
decreases to about $20\%$, because such halos are then growing much
more slowly. When the original semi-analytic code encounters a leaf,
it assumes that it is filled with pristine gas, shock-heated to the
virial temperature. However, hydrogen can cool efficiently at the
temperatures corresponding to minimum mass MXXL halos, so that such
objects should, in fact, already contain galaxies at the first time
they are identified. The absence of these objects explains why
low-mass galaxies (below $\Mstar \sim 10^{11}\,\Mass$) are less
abundant in the MXXL than in the MS. In addition, in halos of any
given (low) mass, MXXL galaxies are typically less massive than MS
galaxies because they have been forming stars for less time.

In order to solve this problem, we have modified the semi-analytic
code so that instead of initialising a ``leaf'' with hot gas alone,
we include a galaxy population taken from a MS halo of similar FoF
mass and redshift. In particular, we randomly chose a MS halo at the
same snapshot and with a tolerance in mass of $\Delta \log_{10} M =
0.1$. (We progressively increase this tolerance up to $2$ when there
are no matches, as is sometimes the case at high redshift for massive
leaves.) The stellar mass function after this modification is shown by
green lines in Fig.~\ref{fig:mf}. It is clear that the increased
abundance of galaxies below $M^*$ brings the galaxy population in the
MXXL much closer to that of the MS. However, a discrepancy persists at
high mass.

This disagreement is again a consequence of mass resolution,
but this time it is caused by the lack of minor mergers onto more massive
haloes in the
MXXL (those events involving accreted haloes in the mass range
$[1.7-12.2]\times10^{10}\,\Mass$, i.e. between $[0.2-20]$
particles).  Although these missing mergers typically contain low
mass galaxies ($\Mstar < 10^{9}\,\Mass$ at $z=0$ in
the MS), they have a noticeable cumulative impact on the most massive galaxies
that our model predicts. In particular, and contrary to a naive
expectation, the net effect of these mergers is not to increase but to
decrease the stellar mass of central galaxies. Although the accreted
subhaloes add stars and gas to the central galaxies, their main effect is
to increase the efficiency of AGN feedback by increasing
the mass of the central supermassive black holes. The more massive a
BH, the more energy it will inject into surrounding gas, thereby
preventing it from cooling and turning into stars.

In order to mimic the effect of unresolved mergers, we have
artificially extended our merger trees by adding branches with mass
below the resolution limit of the MXXL. The number of extra branches
is drawn from a Poisson distribution with mean equal to the average
number of low-mass mergers per halo found in the MS. Typically, the
number of added mergers per halo ranges from $\sim0.02$ to $\sim50$,
for haloes of mass $10^{11}\,\Mass$ to $10^{15}\,\Mass$ at $z=0$,
respectively. We repeat this for every simulation snapshot and in $8$
halo mass bins equally spaced in $\log{M}$. The spatial position of
these new nodes follows a Lorentzian distribution with mean equal to
the virial radius of the main host halo (this is consistent with
direct measurements in the MS).  Note that this modification is
coupled with that discussed before, i.e. these new haloes contain
statistically the correct galaxies.

The resulting stellar mass function is displayed by red lines in
Fig.~\ref{fig:mf}.  This latest modification reduces the abundance of
very massive galaxies, as discussed above, yielding better convergence
between MXXL and MS predictions for galaxies with stellar masses above
$2\times10^{9}\Mass$. In addition, we find a net increase in the
number of galaxies below $M^*$ due to the addition of a population of
low-mass satellites in the range $10^{7} < \Mstar/(\Mass) < 10^{9}$,
which would otherwise be absent in the MXXL. Discrepancies of up to a
few tens of percent still remain at stellar masses $\sim 10^{11}\Mass$
but for the purposes of this paper we regard these as acceptable.

%==================================================================
\begin{figure} 
\includegraphics[width=\linewidth]{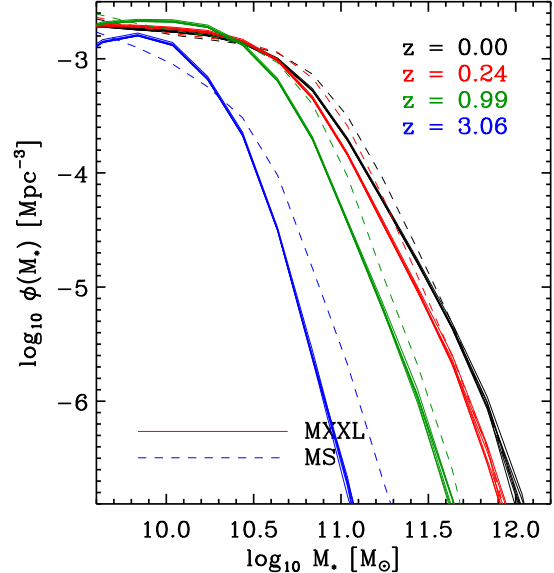} 
\caption{Comparison of the stellar mass functions at different
  redshifts for semi-analytic simulations of galaxy formation in the
  MXXL (solid lines) and MS (dashed lines).
\label{fig:mf_z} } 
\end{figure}
%=================================================================

This shows that, despite the limited resolution of the MXXL, galaxy
formation can be modelled in a way that leads to realistic abundances
at $z=0$ for galaxies with $M_{\rm star}>2\times10^{9}\Mass$. In
Figure~\ref{fig:mf_z}, we compare stellar mass functions in the two
simulations at higher redshift, focusing on the range of stellar
masses well probed by the MXXL. We can see that there is an increasing
mismatch at the high-mass end at higher redshift.  Nevertheless, at
$z<1$ the agreement is of similar quality to that at $z=0$, and at all
redshifts it is reasonable, if one keeps in mind that these numerical
convergence errors are comparable to the observational uncertainties
in stellar mass function estimates and to the systematic differences
between high-redshift observational estimates and the predictions of
the original \cite{Guo2011} model.

The problems discussed above highlight the importance of adequate
resolution in simulations of galaxy formation. Resolution effects can
be significant not only for galaxies in haloes close to the resolution
limit, but also for higher mass galaxies which are substantially
modified through accretion of smaller systems.

%==================================================================
\begin{figure}
\includegraphics[width=\linewidth]{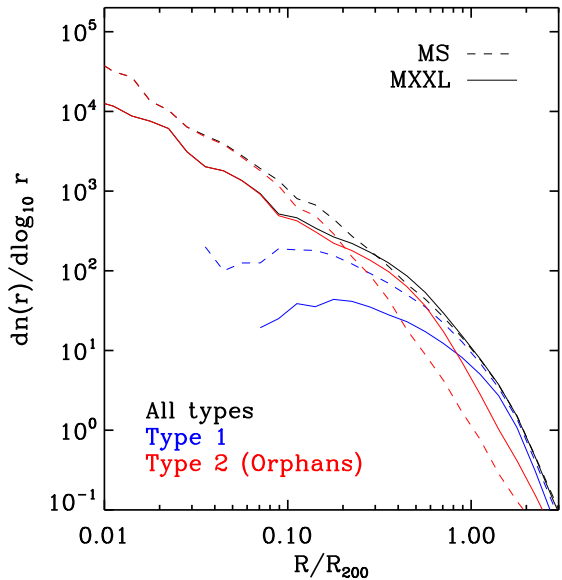}
\caption{Average number density profile for the $40$ satellite galaxies with 
  the largest stellar mass inside clusters in the mass range
  $2\times10^{14} < M/(M_{\odot}) < 5\times10^{14}$. Solid lines show 
  results for the MS, whereas dashed lines show results for the MXXL. Red and blue lines indicate the
  contributions of `type 1' and `type 2' galaxies, those which still
  have an associated subhalo and those which do not, respectively.
\label{fig:profile} }
\end{figure}
%=================================================================

So far, we have addressed artefacts introduced by the differing mass
resolution of the MXXL and the MS. The remaining difference is the
inability of the MXXL to track the position of ``orphan galaxies'',
those whose DM subhalo has fallen below the mass resolution of the
simulation due to tidal stripping. In the MS and MS-II the position of
these galaxies is continuously updated using information about the
position and velocity of the most bound DM particle of the galaxy's
last resolved host halo. This approach was not feasible for the MXXL,
since the full particle data were stored at only a handful of
redshifts in order to avoid an unmanageable data volume (following MS
procedures would have produced $\sim 1$ PetaByte). In the MXXL, once a
galaxy becomes an orphan we assume that its position remains fixed
relative to the central galaxy of its DM halo, except that the
separation linearly shrinks to mimic the effect of dynamical
friction\footnote{We note that this corresponds to the dynamical
friction formula in \cite{Guo2011}, not that in \cite{Guo2011err}}. This has a
negligible impact on the stellar mass function, but dominates the
radial distribution of satellites inside halos, thus affecting
small-scale galaxy clustering.

%==================================================================
\begin{figure}
\includegraphics[width=\linewidth]{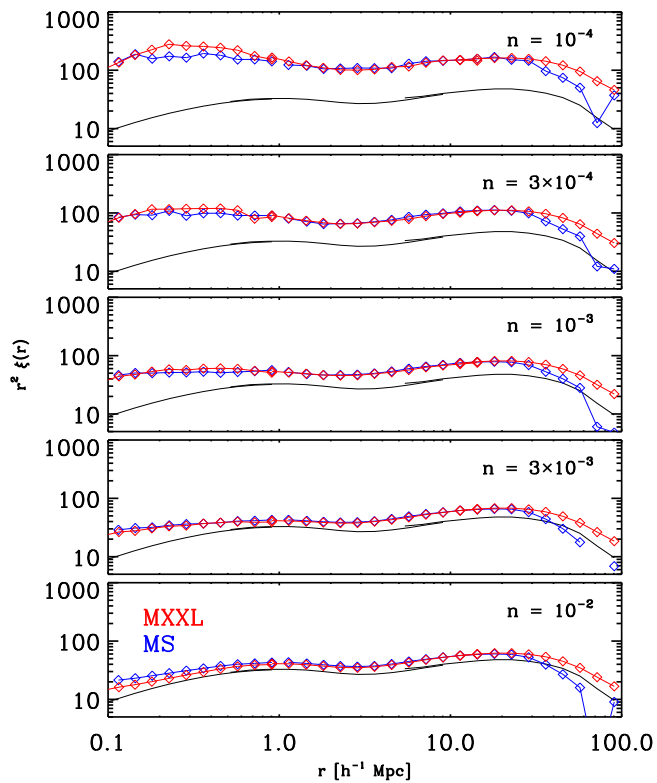}
\caption{Autocorrelation functions at $z=0$ for galaxies in the MXXL
  and MS selected by stellar mass. The five panels show results for
  galaxy samples matching five number densities (given in the legend
  in units of $h^3\,{\rm Mpc^{-3}}$). From bottom to top, they
  correspond to threshold stellar masses of $1.4\times10^{10}$,
  $2.7\times10^{10}$, $5.9\times10^{10}$, $7.3\times10^{10}$ and
  $1.1\times10^{11}$ in units of $\Mass$. Solid black lines denote the
  correlation function of DM in the MXXL.
  \label{fig:xi_test} }
\end{figure}
%=================================================================

To illustrate this, in Fig.~\ref{fig:profile} we compare the average
radial distribution of galaxies inside clusters of mass
$2\times10^{14} < M/(\Mass) < 5\times10^{15}$ in the the MXXL and
MS. The total profiles (solid and dashed black lines) agree very well
down to a scale of about $0.4 h^{-1}{\rm Mpc}$. At smaller scales, MXXL
galaxies display a core which may reflect a misestimation of the
effective orbit and dynamical friction timescale for accreted subhalos.
For this cluster mass, satellites with a resolved DM host are the dominant 
type only at relatively large distances from the cluster center, 
$r > 1 \Mpc$. This is three times larger than the corresponding scale 
for the MS.

To complete this section, we compare 2-point correlation functions
(2pCF) for galaxies in the MXXL and the MS -- the clustering statistic
most relevant for our paper.  This is a demanding test, since it
probes many different aspects of the simulated galaxy population, such
as the way in which galaxies of a given property populate DM haloes,
their correlation with neighbouring haloes, and their radial
distribution within their own haloes.

Here and in the rest of the paper we compute the 2pCF,
$\xi(\vect{r})$, in Fourier space;

\begin{equation}
\xi(\vect{r}) = \mathcal{F}^{-1}\left\{||\mathcal{F}[\delta(\vect{x})]||\right\}
\end{equation}

\noindent where $\mathcal{F}$ and $\mathcal{F}^{-1}$ respectively
denote direct and inverse Fourier transforms, the vertical bars denote
the modulus of a complex field, and $\delta(\vect{x})$ is the input
overdensity field. Operationally, we use FFTs with a mesh of $2048^3$
cells, and a Nearest-Grid-Point assignment scheme. In addition we
repeat this procedure after folding the density field 6 times in each
dimension, thereby enhancing the range of scales probed. This method
has the computational advantages that it is fast, and that only a
subset of the objects need to be in memory at any given time. The
latter is important for analyses of the MXXL due to the large number
of bodies involved.

Fig.~\ref{fig:xi_test} shows the spherically averaged real-space 2pCF
times $r^2$ for galaxy samples selected according to their stellar
mass, for five different number densities, as indicated by the figure
legend. Measurements in the MXXL and in the MS agree remarkably well
over the range $0.4 < r/(\Mpc) < 20$, the regime where both
simulations have adequate volume and resolution. On scales larger than
$20\Mpc$, the MS 2pCFs are systematically lower in all panels than
those found for the MXXL. This is a result of the relatively small box
of the MS, which does not allow correct sampling of the large-scale
modes present in the MXXL. On scales smaller than $0.4\Mpc$,
differences arise due to the poorer spatial resolution of the MXXL and
its reduced ability to track orphan galaxies, as shown in
Fig.~\ref{fig:profile}. In upper panels, the results for the MS are
also somewhat noisy. The excellent agreement on intermediate scales
validates the modifications we have implemented in our semi-analytic
code, and thus the analysis we will present below.

To summarize this section, we conclude that semi-analytic galaxy
formation modelling can be carried out to sufficient accuracy on the
MXXL to allow construction of a realistic galaxy population. In the
next two sections we will investigate further the predicted relation
between the stellar mass of a galaxy and its DM host, as well as, the
appearance of the BAO signal imprinted on the galaxy distribution.

\section{The Galaxy Catalogue}

%==================================================================
\begin{figure*}
\includegraphics[width=0.2\linewidth]{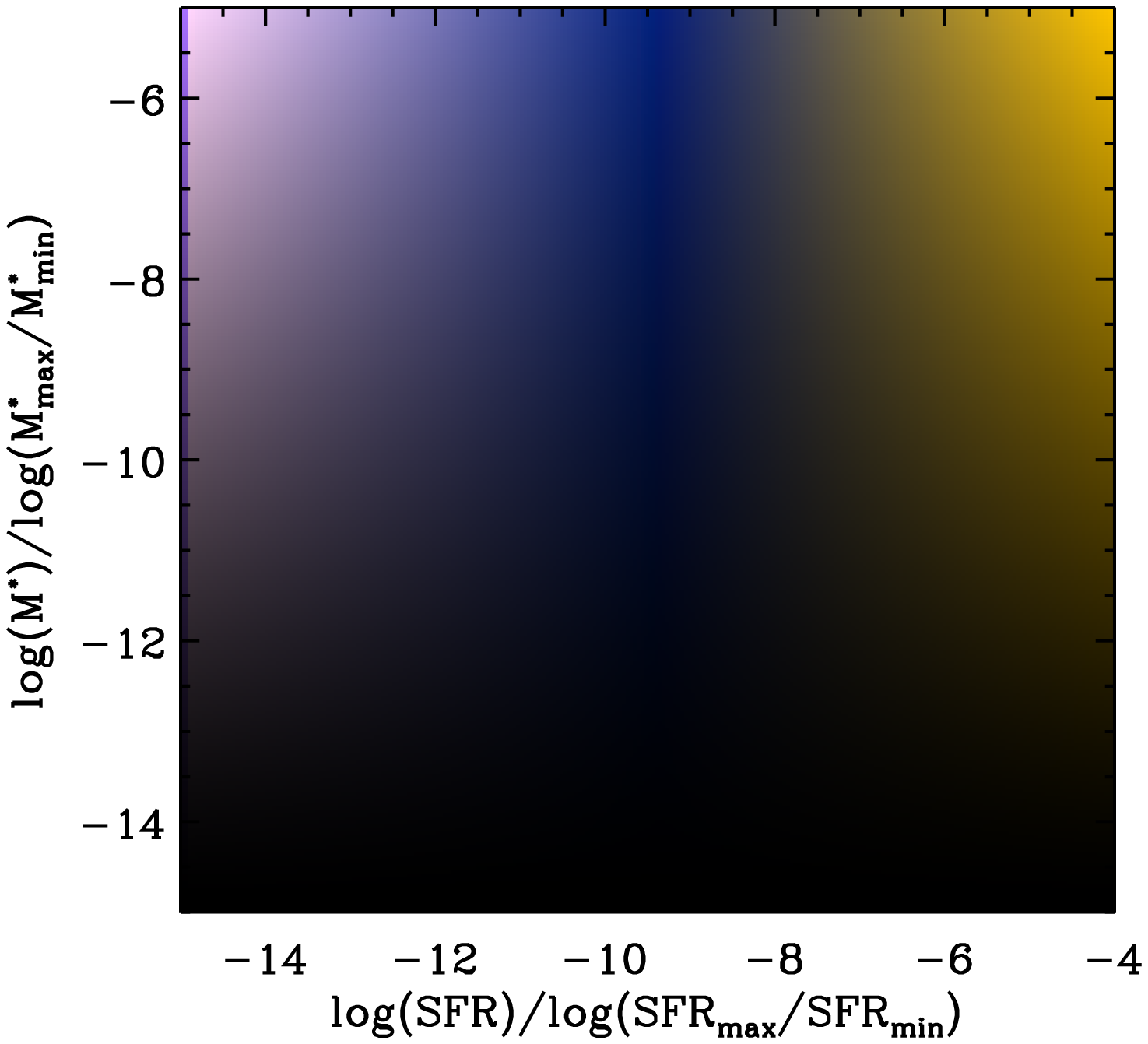}
\includegraphics[width=0.7\linewidth]{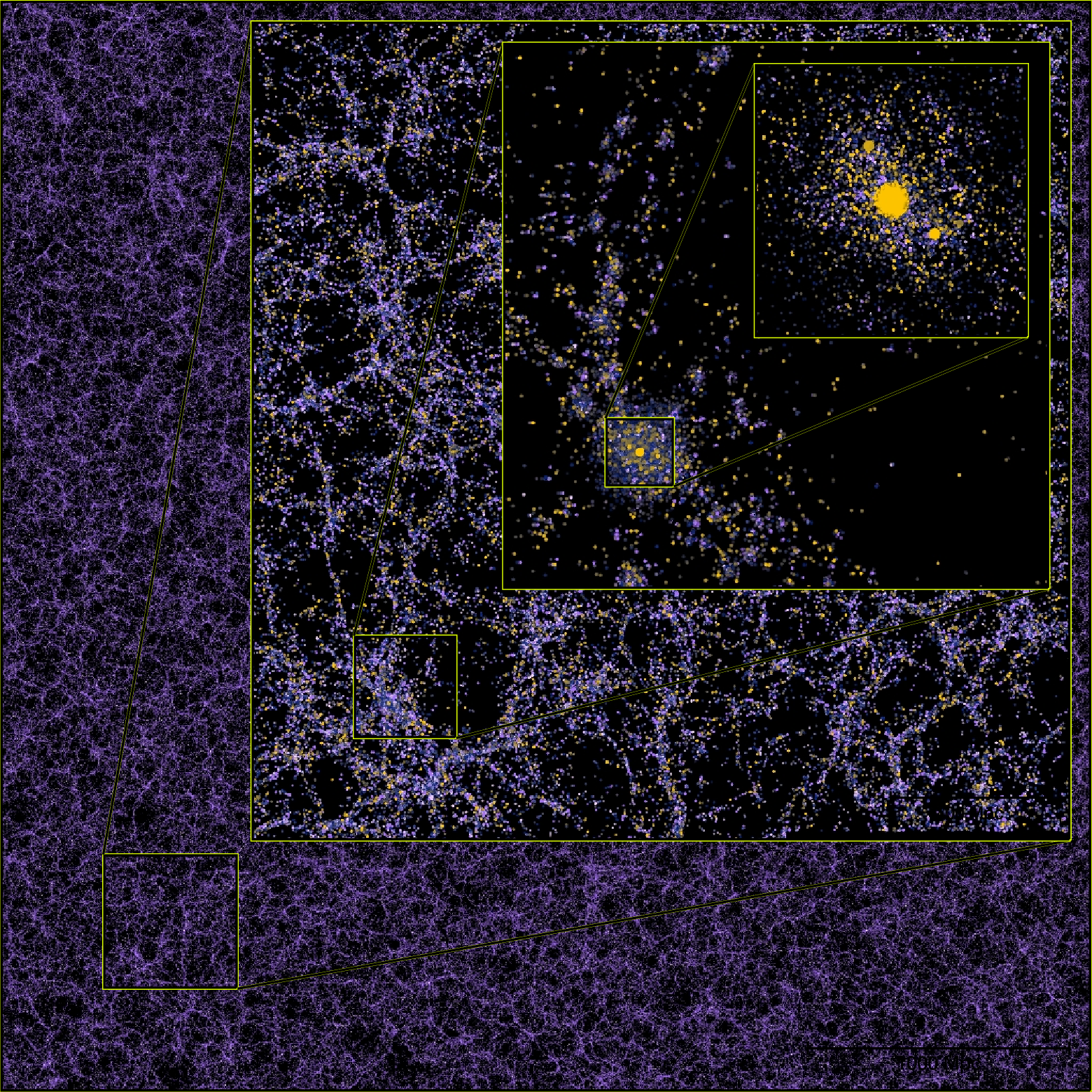}
\caption{The predicted galaxy distribution in a slice of thickness
  13.7 Mpc in the MXXL at $z=0$. Each inset zooms by a factor of $8$
  from the previous one, focusing on the most massive cluster present
  in the simulation. The side-length varies from 4.1~Gpc down to
  8.1~Mpc. Each galaxy is represented by a sphere with intensity and
  size related to total mass in stars and to size of the stellar disk,
  respectively. The intensity in the image is proportional to the
  logarithm of the stellar mass projected along the line-of-sight, and
  the colour encodes the star-formation rate weighted by the stellar
  mass along the line-of-sight.  This simulation has a dynamic range
  of $3\times10^5$ in each spatial dimension, simultaneously resolving
  the internal structure of collapsed objects and the large-scale
  quasi-linear fluctuations expected in a $\Lambda$CDM universe.
\label{fig:picture}}
\end{figure*}
%=================================================================

Within the MXXL volume at $z=0$, our galaxy formation simulation
predicts almost one billion galaxies with stellar mass above
$2\times10^{9}\,\Mass$. For each of them we have followed
both the evolution of its physical properties and
its position and peculiar velocity, thus predicting
galaxy population properties over almost five orders of magnitude
in length, from one hundred kiloparsec up to $4$ Gigaparsec. The full
galaxy formation simulation was carried using $5000$ CPU hours and
produced about $5$Tb of data products.

The large dynamical range covered by our galaxy catalogue can be seen
in Fig.~\ref{fig:picture}, which displays a projection of the
distribution of galaxies at $z=0$ on different scales, zooming
into the most massive cluster in the whole MXXL simulation. The size
and intensity with which we represent each galaxy are set by its disk
diameter and its stellar mass, respectively, while the colour reflects
the galaxy's star formation rate. The 50\% highest star forming galaxies
are depicted in blue, whereas the 50\% lowest star forming are shown in
red. It is evident that the galaxy distribution has similar structure 
to the DM distribution
\cite[see][for a dark matter version of this figure]{Angulo2012b},
showing large clusters, filaments, sheets and voids. It is also,
however, readily apparent that there are marked differences in the way
in which star formation and stellar mass trace those structures and,
in general, the underlying mass density field. This illustrates that
different galaxy properties relate to different aspects of the cosmic
web, and that galaxies do not uniformly sample the underlying DM
field. We now explore quantitatively how this affects measurements of
the large-scale structure of the Universe, in particular the BAO
signal in the 2pCF.

Our main results will be presented in Section~\ref{sec:lss}, but we
begin in Section~\ref{s:mstar} by briefly exploring the relationship
between the properties of galaxies and those of their host DM halos.
This will help us to understand our later results. Then in
Section~\ref{s:gal_catalogues} we will define the galaxy and halo samples
analysed throughout the remainder of the paper.

\subsection{The $M_{\rm halo}-M_{\rm star}-M_{\rm bh}$ relation}
\label{s:mstar}
%==================================================================
\begin{figure} 
\includegraphics[width=8.5cm]{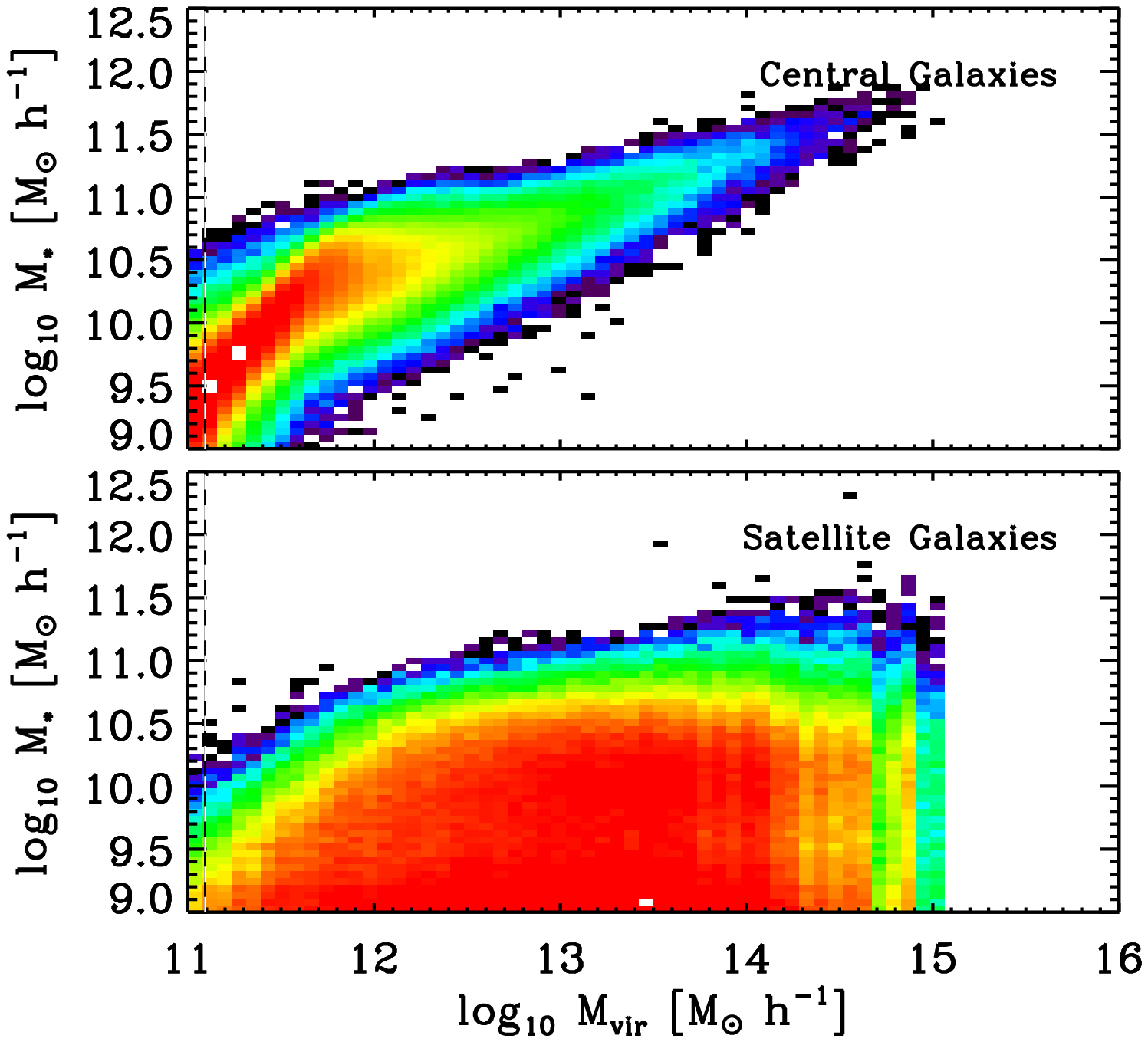} 
\caption{The relation between stellar mass and host halo mass for
  galaxies in our MXXL simulation at $z=0$. For both central and satellite
  galaxies we use the current virial mass $M_{200}$ of their
  halo. The colour in each pixel is set by the logarithm of the number 
  of galaxies in that region of the $M_{\rm halo}-M_{\rm star}$ plane,
   red being the highest density and black the lowest. The vertical 
  dashed line indicates a $20$ particle limit, roughly corresponding 
  to the minimum DM halo mass resolved in our simulation.
\label{fig:MhaloMstar} }
\end{figure}
%=================================================================

We start by showing in Fig.~\ref{fig:MhaloMstar} the abundance of MXXL
galaxies in the plane defined by halo virial mass $M_{200}$ and
stellar mass at $z=0$. In the top panel we focus on central galaxies, and in
the bottom panel we focus on satellites. In both cases, $M_{\rm vir}$
corresponds to the current mass of the host DM halo. 

For central galaxies there is a monotonically increasing and tight
relation between these two quantities. Above $\Mhalo
\sim 5\times10^{11}$, the stellar mass of a galaxy scales roughly as
the halo mass raised to the $1/3$ power with a scatter of $0.2$~dex.
Below $\Mhalo \sim 5\times10^{11}$, the relationship is steeper:
$\Mstar \propto M_{\rm vir}$, with an increasing scatter. We note that
the scatter in {\Mhalo} at a given {\Mstar} is significantly larger;
$0.4$~dex, and also displays a more complex structure.

\begin{figure} 
\includegraphics[width=\linewidth]{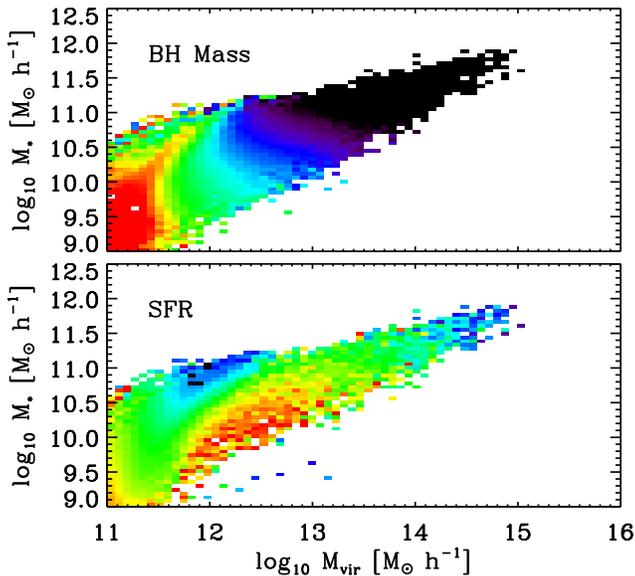}
\caption{The logarithm of the average black hole mass (top) and star
  formation rate (bottom) for central galaxies, as a function of halo
  mass and stellar mass. The red color indicates a BH mass of
  $2\times10^5\,\Mass$, or a SFR of $0.02\,M_\odot\,{\rm yr}^{-1}$,
  whereas black denotes a BH mass of $10^9\,\Mass$ or a SFR of
  $10\,M_\odot\,{\rm yr}^{-1}$, for the top and bottom panels,
  respectively.
\label{fig:BH_SFR} }
\end{figure}

This strong correlation between stellar mass and halo mass reflects
the fact that the mass locked in stars is controlled primarily by the
amount of baryonic material available in the host halo, as modulated
by cooling, star-formation and feedback.  (As we will see in
Fig.~\ref{fig:BH_SFR}, the situation is more complex for the SFR.)
Processes such as feedback are important in that they systematically
change the overall star formation efficiency as a function of halo
mass and redshift, but they only introduce scatter at fixed mass and
redshift, an effect which is particularly marked in the mass range
$10^{11} < M/\Mass < 10^{12}$. This strong correlation is the
principal justification for subhalo abundance matching techniques
\citep[e.g.][]{Vale2004}.

Satellite galaxy masses are much more weakly correlated with host halo
mass, as expected from hierarchical growth. However, although we do
not show it here, they display the same $M_* - M_{\rm halo}$ as
other central galaxies just before they are first accreted onto a more
massive system, and their stellar mass evolves rather little after
this time due to the relatively rapid quenching of star formation in
satellites in our semi-analytic model.

Closer inspection of this figure shows that there is increased scatter
in the central galaxy relation at $\Mhalo \sim 10^{12}\,\Mass$. This
reflects the transition between two modes of galaxy growth that
dominate for small and large halo masses, respectively, and that
overlap at this value of $M_{\rm halo}$ \citep[see][]{Guo2008}.

%==================================================================
%=================================================================

The source of this excess scatter is further clarified in
Fig.~\ref{fig:BH_SFR}, which shows mean black hole mass and mean star
formation rate for central galaxies as a function of their $\Mstar$
and $\Mhalo$. For halo masses between $5\times10^{11}\,\Mass$ and $5\times10^{12}\,\Mass$, deviations from the mean stellar mass, at a fixed $\Mhalo$, correlate with deviations in the
mass of the central black hole and with the star formation rate. This
is because small black holes produce less feedback, and star formation
is set by cooling times and accretion onto these halos. The second
population of galaxies with large black holes reaches similar
star formation rates only for the most massive galaxies.

From these plots, it is clear that the relationship between galaxies
and the underlying DM matter field is not simple. Galaxy properties
relate to their DM haloes in a way that depends not only on current
halo state, but also on halo assembly history -- through, for
instance, the black hole growth history. This is particularly clear
for satellite galaxies, where the galaxy-halo connection is frozen
at the moment of accretion, and galaxy properties relate only weakly
to the current mass either of the galaxy's own subhalo or of its
parent halo. However, this is also the case for a variety of
properties of central galaxies.  For example, at fixed stellar mass,
red central galaxies tend to have more massive haloes than blue ones,
both in the semi-analytic model and in the real world
\citep[e.g.][]{Wang2012}.

\subsection{Definition of our galaxy samples}
\label{s:gal_catalogues}

%=================================================================
\begin{figure} 
\includegraphics[width=\linewidth]{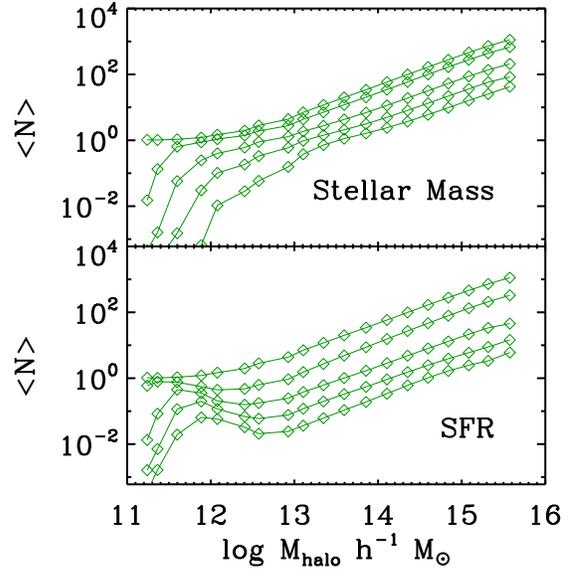}
\caption{The ``halo occupation distribution'' (HOD), the average
  number of galaxies per DM halo as a function of halo mass, for
  galaxies in the MXXL simulation at $z=0$. We show this quantity for
  galaxies selected according to stellar mass (top panel) and
  according to star formation rate (bottom panel). Different curves
  indicate samples at five different number densities. From top to
  bottom, these are $10^{-2}$, $3\times10^{-3}$, $10^{-3}$,
  $3\times10^{-4}$, $10^{-4}\,h^3{\rm Mpc}^3$. Note the different
  shape of these curves in the two cases, especially for samples with
  low number density.
\label{fig:hod}}
\end{figure}
%=================================================================

%==================================================================
\begin{figure} 
\includegraphics[width=\linewidth]{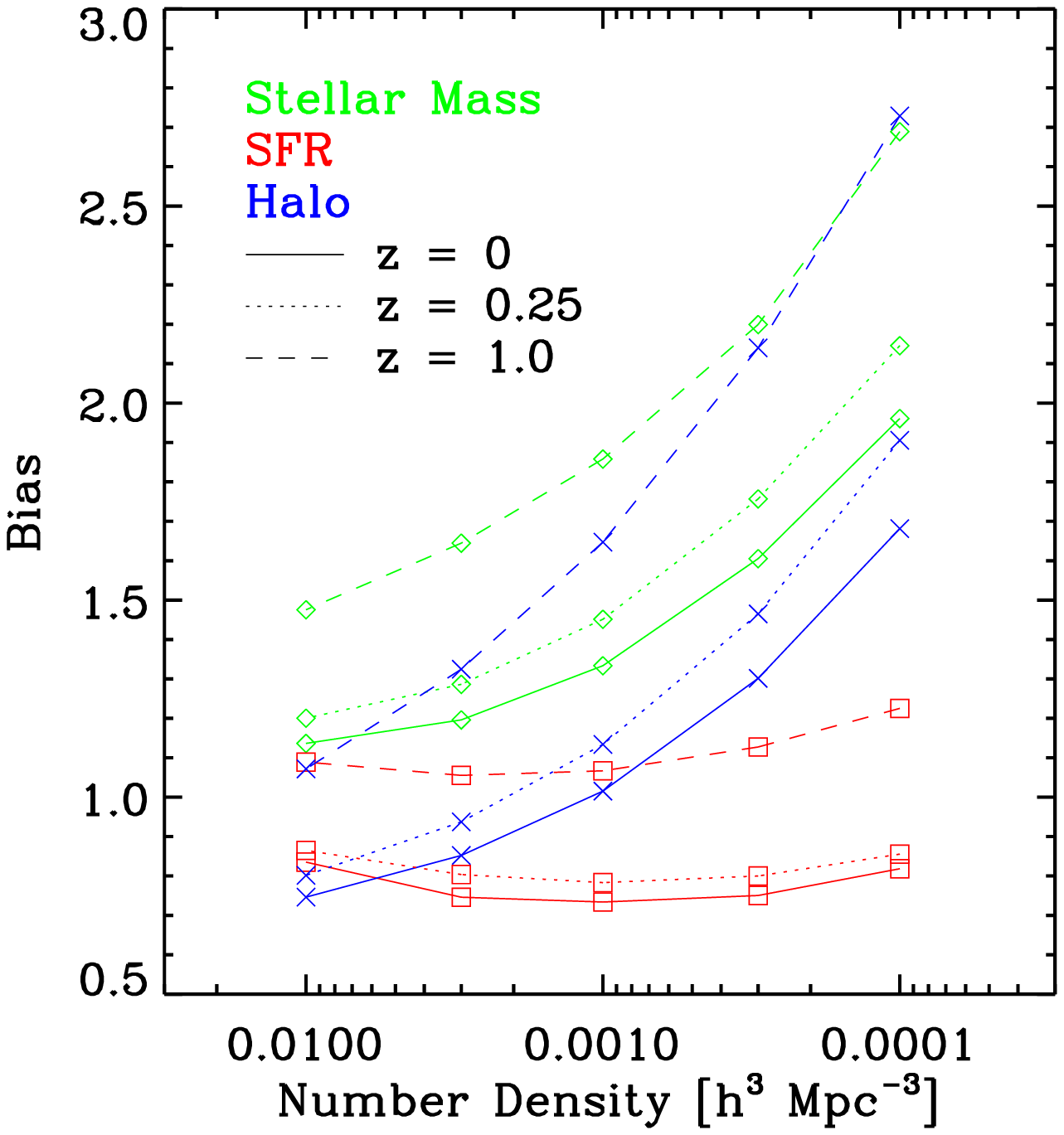} 
\caption{Bias as a function of number density at three redshifts for
  three different samples of objects; dark matter haloes selected
  according to their virial mass, $M_{200}$, galaxies selected
  according to their stellar mass, and galaxies selected according to
  their SFR. The bias is defined here as the square root of the ratio
  of the sample autocorrelation function to the dark matter
  autocorrelation function, averaged over the separation range $60\Mpc
  < r < 70\Mpc$.
\label{fig:bias} } 
\end{figure}
%=================================================================

The complexity of the galaxy-halo relationship results in variations
in the large-scale clustering measured in wide-field galaxy surveys,
because these follow different observational strategies and select
their samples using different criteria. To explore this in our
subsequent analysis, we create two samples of galaxies, one selected
according to stellar mass and the other according to star formation
rate, and for comparison we also select a sample of DM halos according
to mass.

\begin{itemize}
\item[1)] The first galaxy sample is defined using total stellar mass
  (i.e. the sum of the stellar mass of disk and bulge).  As shown
  above, this property correlates quite tightly with halo mass, at
  least for central galaxies. Observationally, a catalogue
  constructed in this way mimics the selection criteria used, for
  example, for the CMASS sample of the BOSS survey
  \citep{Eisenstein2011}.

\item[2)] The second galaxy sample is defined according to
  instantaneous star formation rate (SFR). Since a high star formation
  rate implies a large number of young and massive stars, this
  catalogue mimics selecting galaxies by emission flux, as was done
  for the WiggleZ survey \citep{Drinkwater2010}, as is planned for the
  HETDEX experiment \citep{Hill2008} and as is the current strategy
  for the EUCLID mission \citep{Laureijs2009}. Unlike stellar mass, SFR is not
  expected to track halo mass closely, even for central galaxies,
  since starbursts can be triggered by minor mergers and by disk
  instabilities, and star formation can be quenched by a variety of
  mechanisms.

\item[3)] The third sample consists of halos selected according to
  their virial mass, $M_{200}$. This sample will help us to
  distinguish effects arising from the physics of galaxy formation
  from those that are due solely to nonlinear DM dynamics and halo
  identification.
\end{itemize}

For each of these three definitions, we construct samples of objects
at five different number densities. In units of $h^3 {\rm Mpc^{-3}}$,
these are $[10^{-4},3\times10^{-4}, 10^{-3},3\times10^{-3},10^{-2}]$,
which for the XXL volume corresponds to $2.7$, $8.1$, $27$, $81$ and
$270$ million objects, respectively. For comparison, note that the
CMASS sample of the BOSS survey was designed to have a roughly
constant space density of galaxies, $n = 3\times10^{-4}\,h^3 {\rm
  Mpc^{-3}}$, that the space density of galaxies in WiggleZ is about
$n = 2\times10^{-4}\,h^3 {\rm Mpc^{-3}}$, and that for the planned EUCLID
survey, estimated densities range from $0.15$ to
$4.8\times10^{-3}\,h^3 {\rm Mpc^{-3}}$, depending on the redshift and
galaxy population targeted \citep{diPorto2012}. For J-PAS, the number
density of galaxies with highly-accurate photometric redshift is
expected to range between $10^{-3}$ and $10^{-2}\,h^3 {\rm Mpc^{-3}}$.

The average number of galaxies per DM halo in our ten galaxy samples
is shown in Fig.~\ref{fig:hod}. The top panel displays results for
stellar-mass selected samples, whereas the bottom panel displays
results for SFR-selected samples. Note the non-monotonic behaviour of
the HOD for high-density samples of SFR-selected galaxies. This is
consistent with the previous discussion: the abundance of star-forming
galaxies is not tightly correlated with host halo mass.

In Fig.~\ref{fig:bias} we show clustering bias for all fifteen of our
samples.  Many different effects are visible in this plot. The bias of
stellar mass-selected galaxies increases strongly with decreasing
number density. The slope is almost as steep as for mass-selected DM
haloes, but the overall bias is larger. This behaviour reflects the
strong correlation between stellar mass and host halo mass, together
with the presence of satellites in the galaxy sample. These are almost
always found in haloes more massive than those surrounding central
galaxies of similar stellar mass. The presence of satellites thus
increases the bias. A decrease in the satellite fraction with
decreasing number density and increasing redshift explains the
increasingly similar bias values for stellar mass-selected galaxies
and for DM haloes at higher redshift and lower space density.

Star-forming galaxies show a very different behaviour. Although they
also populate more strongly biased haloes at higher redshift, there is
almost no dependence of bias on the number density of the sample. This
is a consequence of the effect already seen in Fig.~\ref{fig:BH_SFR}.
There is very little tendency for SFR to increase with halo mass. In
addition, as can be seen from Fig.~\ref{fig:hod} the satellite
fraction is significantly smaller in SFR-selected samples than in
stellar mass-selected samples.

\section{Large-scale galaxy clustering}
\label{sec:lss}

%==================================================================
\begin{figure*}
\includegraphics[width=\linewidth]{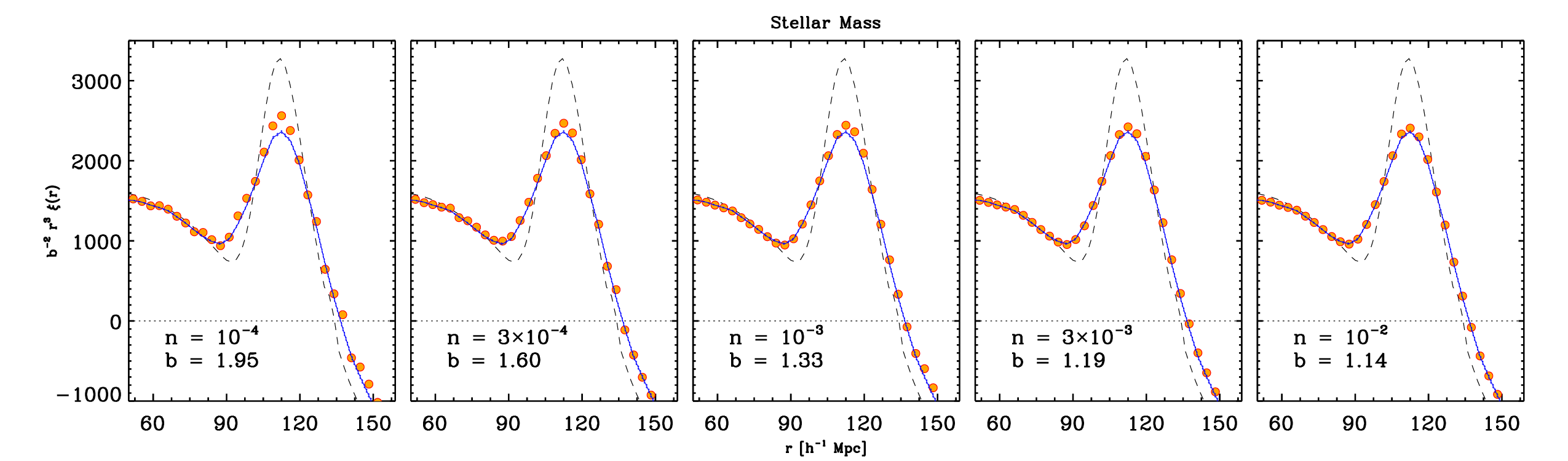}
\includegraphics[width=\linewidth]{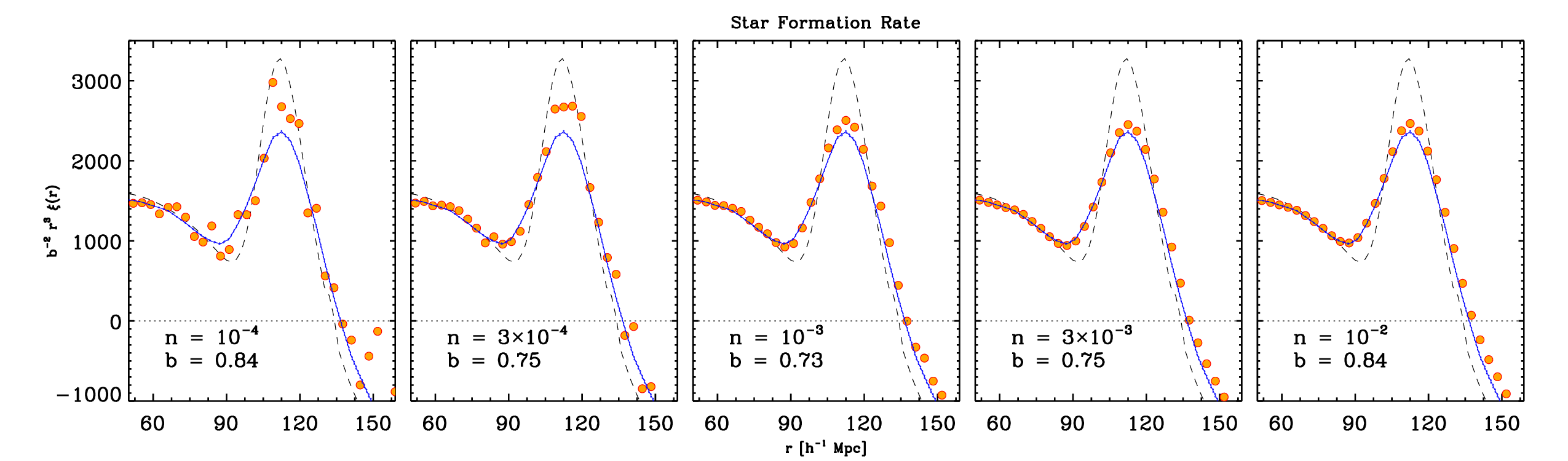}
\includegraphics[width=\linewidth]{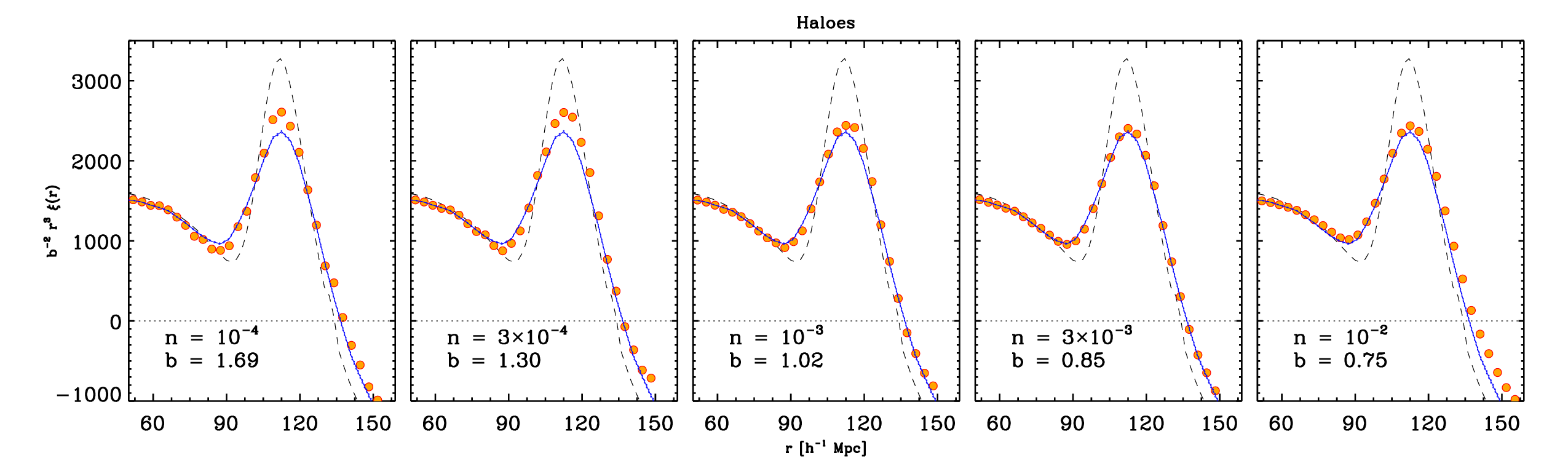}
\caption{The correlation function of galaxies in the MXXL at $z=0$ in
  real space. In the top row, galaxies have been selected according to
  their stellar mass, in the middle row according to their
  instantaneous star formation rate. The bottom row displays a
  catalogue of DM haloes selected according to virial mass
  $M_{200}$. The bias, $b$, and number density of each sample (in
  units of $h\,{\rm Mpc^{-1}}$) is given in the legend. Note that we
  display $\xi(r) \times r^{3} \times b^{-2}$ on the $y$-axis to
  enhance the acoustic peak and to take out the impact of a linear
  bias. For comparison, in each panel we display the 2pCF of the DM
  (blue solid lines) and the linear theory prediction (dashed lines).
\label{fig:bao}}
\end{figure*}
%=================================================================

We now study how the complex relation between the galaxy and DM
distributions may affect BAO measurements from future galaxy surveys

\subsection{The BAO peak in the galaxy correlation function}

Fig.~\ref{fig:bao} shows the $z=0$ autocorrelation functions (2pCF)
predicted by our galaxy formation model for the different samples
described in Section~\ref{s:gal_catalogues}. Filled circles show
results for galaxy samples selected according to stellar mass (top
panel) and according to instantaneous star formation rate (middle
panel). For comparison, we also show results for mass-selected halo
samples (bottom panel). Each column corresponds a different number
density, as indicated by the legend.

Note that we display $r^3\times\xi(r)$ in order to focus on the BAO
signal, which appears at $r\sim110\Mpc$ for our choice of cosmological
parameters. In addition, in this plot we have taken out the effects of
a scale-independent bias by dividing the $y$-axis by the ratio between
the galaxy and DM 2pCF's, averaged over the range $60 < r/\Mpc <
70$. Finally, for comparison we show the $z=0$ linear theory
prediction for the 2pCF (dashed line) as well as the actual $z=0$ 2pCF
for the DM in the full MXXL (blue line).

In all galaxy samples we can identify the acoustic peak at high
signal-to-noise.  The large volume sampled by the MXXL,
$V=27(\Gpc)^3$, results in very small cosmic variance errors, and the
remaining statistical fluctuations arise primarily from shot noise
(i.e. Poisson noise reflecting the finite number of objects in our
samples). For this reason, the scatter in our measurements decreases
from left to right, and it is larger in the top panels than in the
middle ones. At a fixed number density, SFR-selected galaxies cluster
considerably more weakly than stellar-mass-selected galaxies
(c.f. Fig.~\ref{fig:bias}), so that the signal-to-noise is lower and
the data points display larger random fluctuations.

All measured 2pCFs differ substantially from linear theory
expectations (the dashed lines), even on these large scales. This
effect has been discussed by many authors for the DM and halo
distributions. It reflects nonlinear coupling of (initially
independent) Fourier modes, which smears out the BAO peak
\citep[e.g.][]{Seo2003,Angulo2005,Angulo2008,Crocce2008}. Here, we
show that this effect is also visible in the galaxy distribution and
that, to lowest order, it has the same magnitude in all samples,
independent of their bias or number density.  However, upon closer
inspection, we see that significant differences are present in the
amplitude of the BAO peak. We investigate this further in the next
figure.

%==================================================================
\begin{figure*}
\includegraphics[width=\linewidth]{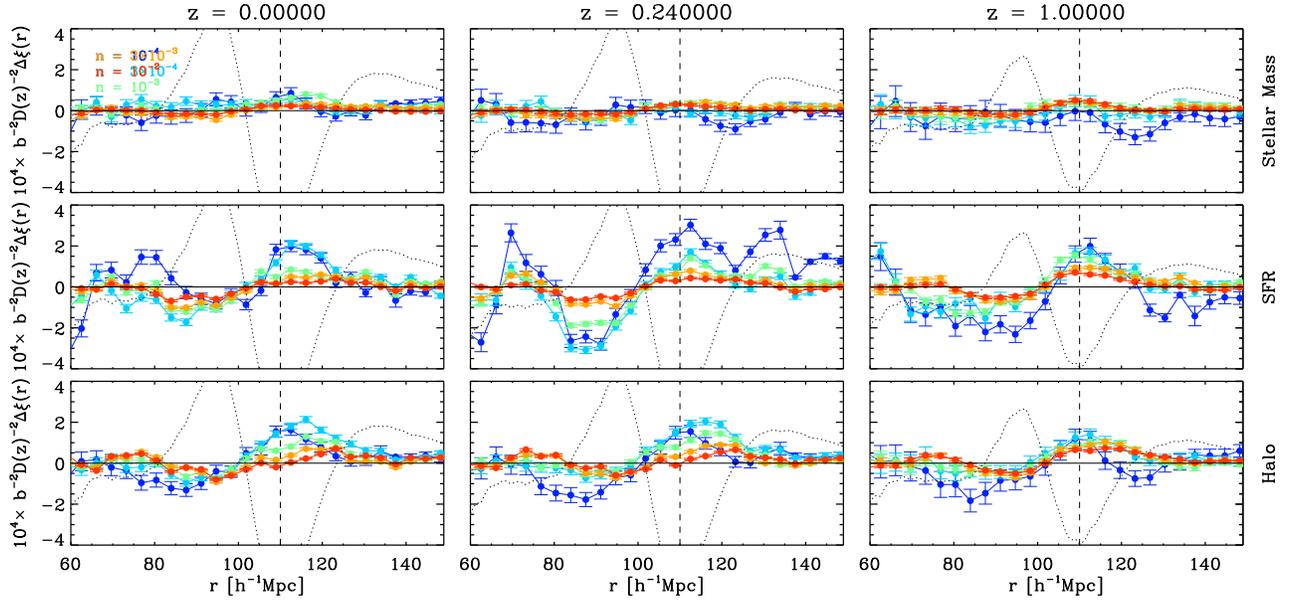}
\caption{Scale dependence of the bias for galaxy samples selected
  according to stellar mass (top row) and to star formation rate
  (middle row), as well as for a mass-selected halo sample (bottom
  row). Coloured lines show results for samples at five different
  number densities, as indicated by the legend, matching the samples
  shown Fig.\ref{fig:bao}.  Vertical dashed lines denote the position
  of the BAO peak in the correlation function of dark matter.
  Deviations from zero imply deviations from linear biasing of the
  nonlinear dark matter distribution. The error bars are given by the
  square root of the diagonal elements of the covariance matrix for
  each measurement. 
  \label{fig:bao_diff_0} }
\end{figure*}
%=================================================================

%==================================================================
\begin{figure*}
\includegraphics[width=\linewidth]{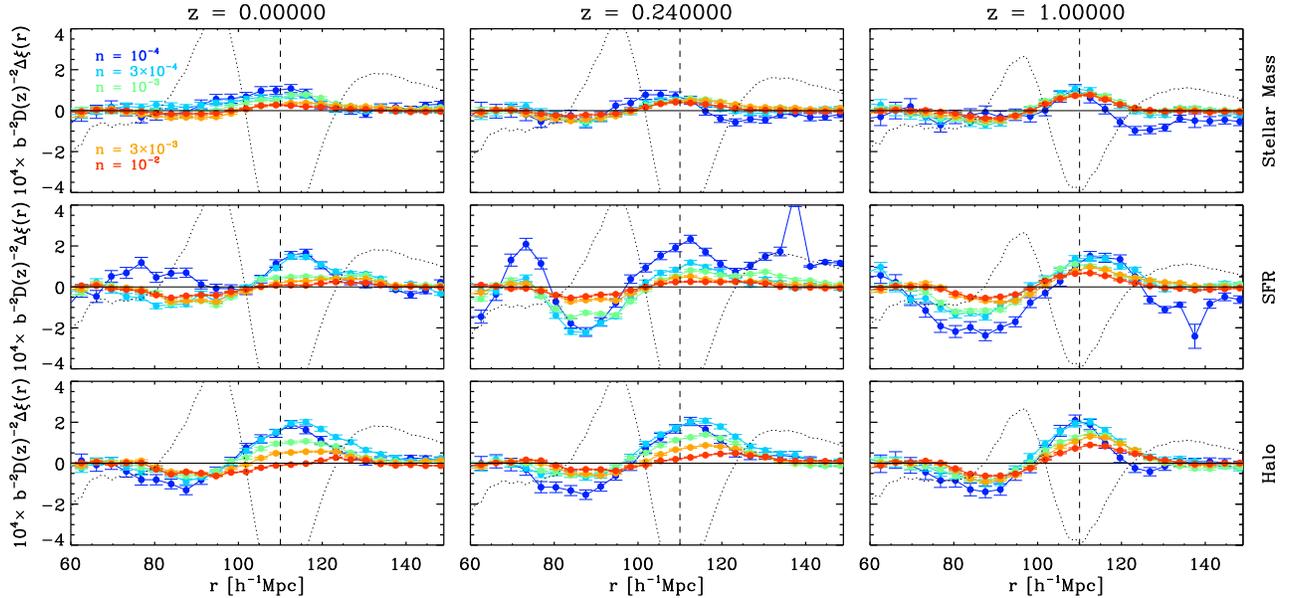}
\caption{Same as Fig.~\ref{fig:bao_diff_0} but for the monopole of the
  redshift-space correlation function.
  \label{fig:bao_diff_1} }
\end{figure*}
%=================================================================

Fig.~\ref{fig:bao_diff_0} displays the difference between the 2pCF of each
galaxy sample and that of a linearly biased version of the DM 2pCF:
$\Delta\xi = \xi_{g,g} - b_0 \xi_{m,m} = [ b(r)^2 - 1 ]\,\xi_{m,m}$, 
with $b(r) = \xi_{\rm g,m}/\xi_{\rm m,m}$, where
$\xi_{\rm m,m}(r)$ is the DM autocorrelation and $ \xi_{\rm g,m}(r)$
is the cross-correlation between DM and galaxies. \footnote{We have explicitly checked that on large
scales, the shape and amplitude of the scale dependent bias derived
from cross- and auto-correlation functions agree extremely well.}
$b_0$ is set to the
average value of $b(r)$ in the range $r=[60-70]\Mpc$. We display 
$10^{4}\,b_0^{-2}\,D(z)^{-2}\,\Delta\xi$ in order to facilitate the comparison
across different redshifts and bias values. Each panel
focuses on a different combination of redshift and selection
criterion. Within each panel, lines with different colours show
results for samples of different density, as indicated by the
legend. We highlight the position of the BAO peak using a vertical
dashed line. Error bars are set by the diagonal elements of the
cross-correlation covariance matrix, which we compute analytically
following \cite{Smith2009}, including the effects of finite volume and
finite tracer number. For comparison, we also overplot the difference
between the linear and nonlinear DM 2pCF.

We highlight two facts that allow us to explore our results with high
precision. By defining $b(r)$ using the measured DM 2pCF at the
relevant redshift, our results are essentially cosmic-variance-free.
Further, the use of cross-correlations (instead of autocorrelations)
greatly suppresses the impact of shot-noise in our results
\citep{Gao2007,Angulo2008b,Smith2009}. This is thanks to the large
number of simulation particles in the MXXL. We note that some residual
noise is still present even in the Gaussian case \citep{Smith2009}.

We recall that if our samples were simply linearly biased versions of
the underlying DM field, then all curves would lie on top of the
horizontal line, i.e. they would be equal to zero on all
scales. It is clear, however, that this is not true, and systematic
deviations from zero appear at many scales, most prominently at the
location of the BAO peak.  Quantitatively, our measurements are
inconsistent with the linear bias hypothesis at the $5-20\sigma$
level, as computed using the full covariance matrices. The deviations
show a similar structure in most samples: by construction, they are
consistent with zero at $r \sim 60-70\Mpc$. They show a suppression
at $r \sim 90$\Mpc followed by an excess at $r \sim
110$\Mpc (the BAO peak) of slightly higher amplitude. On larger scales, they
seem to approach zero again. The deviations appear to show systematic
differences in shape and amplitude between the three different kinds
of sample. In particular, the deviations appear to have lower
amplitude for the mass-selected galaxies than for the SFR-selected
galaxies or the mass-selected haloes. Also, at $z=0$ and $z=0.24$,
there is a trend with halo mass, where a lower mass threshold (i.e.
a higher number density) results in a smaller deviation from zero,
compared to more massive and less abundant halos.

In Fig.~\ref{fig:bao_diff_1} we show an analogous plot but this time
focusing on the monopole of the redshift-space 2pCF.  For this, we
have employed a plane-parallel approximation, and the position along
the $z$-direction includes the contribution of peculiar
velocities. Thus, this is closer to the observed galaxy correlation
function. As in the previous plot, we display results using the
cross-correlation of the galaxy and dark matter field, both measured
in redshift space. Here, there are also deviations from the linear
bias model, indicating scale-dependent bias at the BAO position. These
deviations show a structure and amplitude consistent with those in
real space. In both cases, the net effect of galaxy formation appears
to be an enhancement of the BAO peak. We note that the deviations are
opposite to those caused by nonlinear evolution or RSD, which decrease
the contrast of the BAO peak.

%=================================================================
\begin{figure} 
\includegraphics[width=\linewidth]{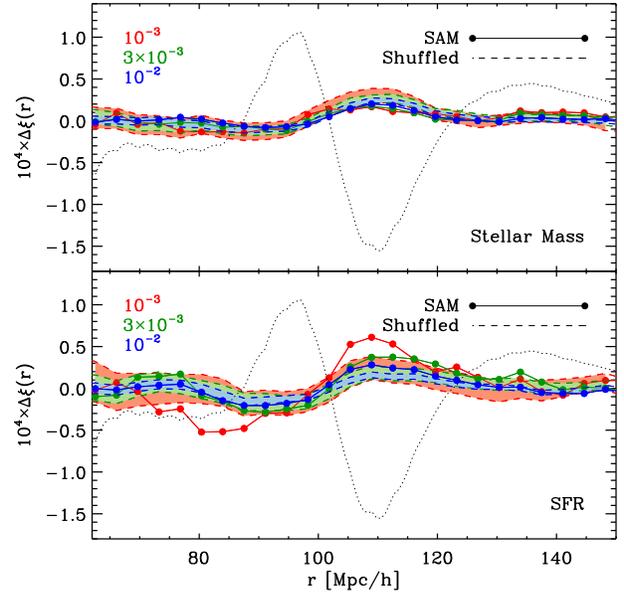}
\caption{The scale-dependent distortions in the large-scale clustering
  of galaxies at $z=1$, for two models of galaxy formation and for
  three samples of different number density. Solid lines show results
  from our original semi-analytic galaxy formation model (SAM),
  whereas dashed lines show results for a Halo Occupation Distribution
  (HOD) model built by shuffling the SAM galaxies among halos at fixed
  halo mass.  Filled regions show the {\it rms} scatter among 100
  realisations of the shuffled catalogue. \label{fig:shuffled} }
\end{figure}
%=================================================================

\subsection{Scale-dependent bias of galaxies and haloes}

As we have discussed before, the galaxies in our model relate in a
nontrivial way to the DM field. In particular, galaxy properties
depend not only on halo mass but also on the details of halo accretion
and merger histories, which in turn are related to halo environment.
There are many effects that could produce the BAO distortions seen
previously. As a test, we have "shuffled" the galaxy catalogues as in
\cite{Croton2007}: we randomly reassign the galaxy populations among
halos of a given mass by taking all the galaxies from halo A
(including the central one) and putting them in halo B using the
original halo B central galaxy position and velocity, but the
properties and position/velocity offsets from halo A. This procedure
guarantees that the HOD of the shuffled catalogue is identical to that of the
original simulation (including all count moments, statistical fluctuations, 
internal spatial and velocity distributions and deviations from
sphericity/isotropy) but it eliminates any spatial correlation between
the populations of disjoint halos.

We have created $100$ of these shuffled catalogues to assess the
associated noise. For each of them we measured the 2pCFs and computed
the deviations from a linear bias model in the same way as in our
original catalogues (shown in Figs.~\ref{fig:bao_diff_0} and
\ref{fig:bao_diff_1}). If the clustering of galaxies is adequately
described by an HOD model based on halo mass alone, then the
scale-dependent bias found for the original simulation should be
statistically consistent with that found for the ensemble of shuffled
catalogues. A systematic difference would imply that additional
variables than halo mass are important.

Fig.~\ref{fig:shuffled} shows our results in real space. For clarity,
we display only the three highest number densities and restrict
ourselves to $z=1$. At other redshifts and at lower number density,
the noise in the measurements is too large for robust conclusions.
The upper and lower panels show results for galaxy samples selected by
stellar mass and by SFR respectively.  Filled circles show the
measurements for the original catalogues, whereas shaded regions show
the {\it rms} scatter among the results for the 100 shuffled
catalogues. Although not shown here, we have checked that very
similar results are found in redshift space.

In all samples, the deviations from zero have similar shape in the
original and in the shuffled catalogues. The disagreements are nowhere
statistically significant when galaxies are selected by stellar
mass. Interestingly, however, a significant effect does appear to be
present when galaxies are selected by star formation rate, although it
is quite weak at high sample densities.  To the accuracy of our
present results, the distortion of the BAO peak appears to be adequately
accounted for by an appropriate HOD model, with the possible exception of
SFR-selected samples of relatively low density. We note that
\cite{Eisenstein2009}, \cite{Mehta2011}, \cite{Wang2013} and \cite{Angulo2013d}
have all reported detections of scale-dependent bias at the BAO location for
halo catalogues analogous to the one analysed here, and indeed such distortions
are expected from theoretical arguments based on the peak-background split
formalism \citep{Desjacques2008,Desjacques2010}.

Although galaxy formation physics appears not to introduce further
distortions in the BAO peak beyond those already present in the halo
distribution, astrophysical processes do affect the peak by setting
the HOD. Incorporating these effects directly in modelling of the BAO
peak would require {\it a priori} knowledge of the HOD of the target
galaxy sample, and of its evolution with redshift. This is because,
as we have shown above, the deviations from a linear bias are different
for halos of different mass. In principle, this
information can be extracted from the observational data, but this
will result in weakening, and possibly in biasing, of the constraints
on cosmological parameters. We will explore this in the next
subsection.

\subsection{Biases in cosmological parameter constraints}

%==================================================================
\begin{figure}
\includegraphics[width=\linewidth]{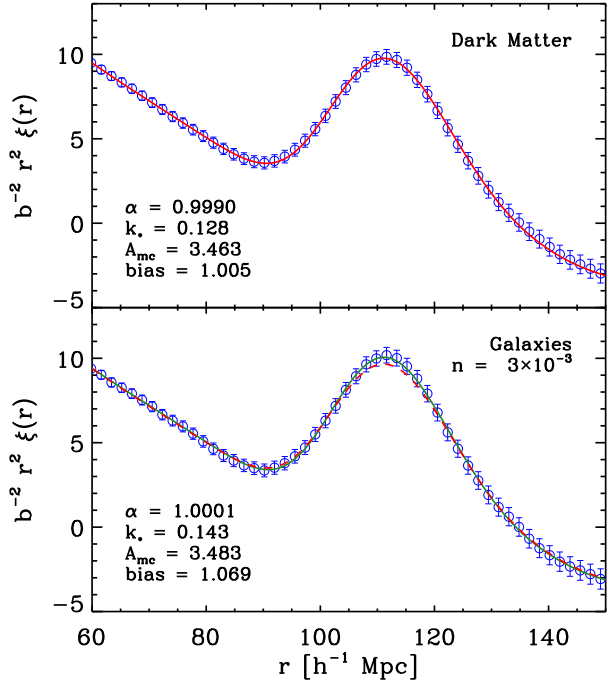}
\caption{ Correlation functions for DM (top panel) and for a
  SFR-selected galaxy sample (bottom panel), together with the best
  fit model (solid lines) provided by Eq.~(\ref{eq:model}).  Error
  bars are given by the square root of the diagonal elements of the
  respective covariance matrices. The DM correlation is repeated in
  the lower panel as a dashed line.
  \label{fig:fit} }
\end{figure}
%=================================================================

We now quantify the impact of scale-dependent galaxy bias on
future cosmological constraints. For this, we have fitted a four
parameter model (similar to those applied in the analysis of real
data) to our correlation function measurements. We follow
\cite{Sanchez2008}, who proposed a model for the shape of the galaxy
correlation based on renormalised perturbation theory
\citep{Crocce2008}. This model has been applied to real data several
times \citep[e.g.][]{Sanchez2009,Sanchez2012}, and it reads as
following:

\begin{equation}
\label{eq:model}
\hat{\xi_g}(r) = b^{2} [\, \xi_L(\alpha r) \otimes {\rm e}^{-(k_* r)^2} + 
 A_{\rm MC} \, \xi'_L(\alpha r)\xi_L^{(1)}( \alpha r) \,] 
\end{equation}

\noindent where $\otimes$ denotes a convolution, $\xi'_L$ is the 
derivative of the linear theory correlation function, and $\xi_L^{(1)}$ 
is defined as:

\begin{equation}
\xi_L^{(1)} = 4 \pi \int P_L(k) j_1(k r) k \, {\rm d}k , 
\end{equation}

\noindent where $j_1(k r)$ is the spherical Bessel function of the
first kind.  The free parameters of the model are ($b$, $k_*$, $A_{\rm
  MC}$, $\alpha$) where $b$ is the sample bias, $k_*$ and $A_{\rm MC}$
control the amount of nonlinear evolution, and $\alpha$ is a
``stretch'' parameter that accounts for possible shifts in the BAO
peak position due to effects not modelled, for example,
scale-dependent bias. If $\alpha$ is unity, the fitting provides
unbiased measurements of the BAO location and thus of the associated
distance scale. If $\alpha \neq 1$, then biased distances and 
cosmological constraints will result. We note that similar approaches
based on a ``stretch'' parameter have been employed in the past to
quantify the cosmological biases introduced by nonlinear evolution and
redshift-space distortions \citep[e.g.][]{Seo2007, Angulo2008,
  Sanchez2008}.

We find the set of best fit parameters, $(\alpha,k_*,A_{\rm MC},b)$,
using a Monte Carlo Markov Chain (MCMC) algorithm (with 20,000 steps) 
that minimises a $\chi^2$ function over separations $r=[70-140]\Mpc$:

\begin{equation}
\chi^2 = \left[\xi_g(\mathbf{r}) -
  \hat{\xi_g}(\mathbf{\mathbf{r}})\right]^{T}
C^{-1}(\mathbf{r},\mathbf{r'}) \left[\xi_g(\mathbf{r'}) -
  \hat{\xi_g}(\mathbf{r'})\right] ,
\end{equation}

\noindent where $C^{-1}(\mathbf{r},\mathbf{r'})$ is the inverse of the
relevant covariance matrix. We estimate $C$ analytically as described
in \cite{Sanchez2008}, including the effects of cosmic variance,
shot-noise, galaxy bias, and correlation function binning. As
discussed by \cite{Sanchez2008}, this approach closely agrees with the
covariance matrix measured from simulations. Finally, we note that for
the input linear theory correlation function, $\xi_L$, we have used a
$z=10$ measurement of the actual DM field in the MXXL simulation. In
this way, we minimize the impact of cosmic variance on our best fit
values. In addition, we note that predictions for galaxy 2pCF's were
constructed by cross-correlating simulated galaxies with DM particles
in the MXXL. In this way, we minimize the impact of shot noise in our results.

%==================================================================
\begin{figure}
\includegraphics[width=\linewidth]{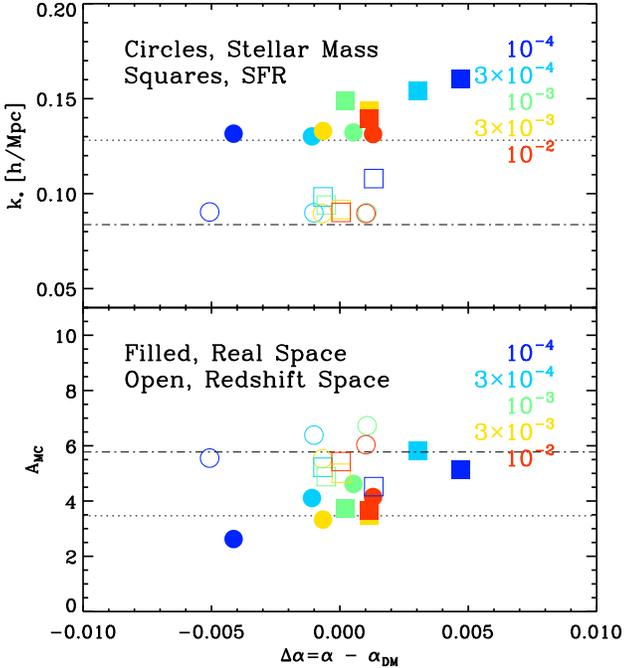}
\caption{The best fit values for $\Delta\alpha$ and $k_*$, and for
  $\Delta\alpha$ and $A_{\rm MC}$ for various simulated galaxy
  samples. Colours indicate samples of different number density. Open
  and filled symbols indicate results in real and redshift space,
  respectively. Squares and circles indicate selection by SFR and by
  stellar mass, respectively. Note the $x$-axis displays the
  difference with respect to the DM best fit value. Horizontal dotted (dot dashed)
  lines indicate the best fit $k_*$ and $A_{\rm MC}$ values for the DM
  correlation functions in real (redshift) space.
  \label{fig:fit_par} }
\end{figure}
%=================================================================

In Fig.~\ref{fig:fit} we show the quality of the fit for two cases at
$z=1$ in redshift space. The top panel shows the 2pCF for DM and the
bottom the 2pCF for a SFR-selected galaxy sample with a number density
$n=3\times10^{-4}\,h^{-3}\,{\rm Mpc^{-3}}$. We recall that the latter can be
regarded as representing the expected results from the EUCLID
satellite. Error bars correspond to the diagonal elements of the
relevant covariance matrix.  The best fit model is shown by solid
lines, and the best fit parameters appear in the legend. For
comparison, in the bottom panel we show the best fit for the DM as a
dashed line.

In both cases, the model provides an excellent description of the
data, resulting in very low values for the reduced $\chi^2$. However,
the best fit itself is different in the two cases, as can be seen by
comparing the solid and dashed lines in the bottom panel.  This
disagreement could lead to bias in cosmological constraints, or it
could be absorbed by another parameter of the model. The latter is
indeed a reasonable expectation. Since galaxy formation effects appear
with similar strength (but opposite sign) to nonlinear evolution and
RSD effects, the model may absorb them by assigning slightly larger
values to the $k_*$ parameter. We explore this next.

In Fig.~\ref{fig:fit_par} we compile the best fit parameters $\alpha$
and $k_*$ (top panel); and $\alpha$ and $A_{\rm MC}$ (bottom panel),
for all the galaxy samples we have considered so far: selected
according to SFR (circles) or stellar mass (squares), and in real
(open symbols) or redshift space (filled symbols). We focus on $z=1$
since it is the target redshift of future large-scale surveys. The
$x$-axis shows the differences with respect to the best fit $\alpha$
parameter for the DM autocorrelations ($0.99903\pm0.0028$ and
$1.0004\pm0.0035$ for the real and redshift space cases, respectively).
By displaying $\Delta\alpha$, we take out deviations from unity
resulting from residual cosmic variance and from shortcomings of the model when
describing the nonlinear DM correlation function. The best fit DM
$k_*$ and $A_{\rm MC}$ values are indicated by the horizontal dotted
lines.

As expected, values of $k_*$ for redshift space are smaller than those
for real space. Nonlinear RSD further weaken correlations, washing out
the BAO peak and inducing a damping term which affects larger
scales. This is seen both for DM and for our galaxy samples,
independent of their number density. Values for $A_{\rm MC}$, which
are mostly constrained by small separations, range from $3$ to $7$
and do not differ significantly between real and redshift space.
These best fit values are consistent with previous studies
\citep{Sanchez2008}.

The deviations of best fit values for $\alpha$ with respect to the 
DM case is for the three densest galaxy samples smaller than $\pm 0.2\%$, 
and smaller than $0.5\%$ for the two lowest
density samples. This is below the target accuracy of most future surveys. The 
deviations from the unbiased
case, $\Delta\alpha = 0$, show almost no correlation with $A_{\rm MC}$,
or $k_*$, and they also seem to be independent of the galaxy sample considered.
For the lowest density sample, the deviations are negative when the stellar
mass is used to select galaxies, and positive when the SFR is used,
which suggest that it has a statistical origin.  
Note that the deviations from a linear bias we observe at $z=1$ are typically
larger than those observed at lower redshifts.

As speculated above, scale-dependent biases can indeed be absorbed by
the free parameters of the fitting model. In particular, in all cases, and
in both real and redshift space, the best
fit value for $k_*$ is slightly larger than that measured for the DM
field. This is consistent with the nature of the scale-dependent bias
we measured in the previous section: the net effect is to enhance the
contrast of the BAO peak, which is the opposite of nonlinear
broadening of the peak, and thus a larger value for $k_*$ is
needed. These are encouraging results for future missions, in
particular, for the galaxy sample matching that planned for EUCLID, we
detect deviations of $0.1\%$, confirming that current modelling
techniques are sufficiently accurate and flexible to properly exploit
future BAO measurements. Nevertheless, propositions to employ the 
BAO damping scale to constrain cosmological parameters or modified
gravity theories \citep[e.g.][]{Cervantes2012} could be seriously 
limited by the galaxy formation effects discussed here.

\section{Conclusions}

The measurement of distances in the Universe via the BAO peak is
currently one the most promising ways of constraining the cosmic
expansion history. This is, in part, thanks to the theoretical
understanding, quantification and correct modelling of nonlinear
evolution and redshift-space distortions, effects that modify the
appearance of the BAO peak in the galaxy distribution.

In this paper we have explored the impact of galaxy formation physics
on the detectability and modelling of the BAO peak. Our approach
followed self-consistently the evolution of over a billion galaxies in
a $70\,{\rm Gpc}^3$ volume. We detect at high significance a
scale-dependent bias at the BAO location, which has the net effect of
enhancing the BAO contrast. We showed that the main agent causing this
was already present in the halo distribution, with galaxy formation
adding negligible effects for stellar mass selected samples and small
effects for SFR-selected samples. Although galaxy formation physics
does not add a significant extra effect, it does enter the problem
indirectly by setting the halo occupation distribution, and thus the
effective magnitude for the scale-dependent bias.

Although current models to extract cosmological constraints from the
BAO were not designed to account explicitly for such scale-dependent
biases, we show that they do indeed have enough flexibility to absorb
these distortions in other nuisance parameters and to return
measurements for the BAO location which are unbiased at the $0.2\%$
level for most galaxy samples at $z=0$. This is fortunate news for 
future galaxy surveys and confirms the robustness of the BAO peak and 
its potential as a standard ruler.

We note that our galaxy formation modelling is still simplified in
many respects, and our conclusions are valid only within our specific
model of galaxy formation.  Although this is one of the most
sophisticated and realistic prescriptions that can currently be
applied to DM-only simulations, there are a number of assumptions and
simplifications that may not hold in reality. In particular, the
modifications to the semi-analytic model described in Section~2.2 will
reduce correlations between galaxy properties and environment. Another
limitation is that our modelling technique forces us to neglect the
back-reaction that baryons exert on the DM distribution. Similarly, we
have assumed that the density evolution of baryons and DM is
identical, which is certainly not true in detail \citep{Angulo2013a,
  vanDaalen2013}.  All these effects are expected to be smaller than
the ones discussed in this paper, but as the precision of measurements
and our physical understanding of galaxy formation continue to
improve, we may expect that yet larger and more sophisticated
simulations will be needed to support robust and precise estimation of
cosmological parameters from galaxy surveys.

\section*{Acknowledgements}

We would like to thank Ariel Sanchez for several useful
suggestions and comments on the manuscript. Financial support from
the Deutsche Forschungsgemeinschaft through Transregio 33, ``The Dark
Universe'', is acknowledged. RA, BH and SW acknowledge support from ERC
Advanced Grant 246797 ``GALFORMOD''.

%----------------------------------------------
%\bibliographystyle{mn2e} \bibliography{gal}
\bibliographystyle{mn2e} \bibliography{database}
%---------------------------------------------------------------------

\label{lastpage} \end{document}